\documentclass[letterpaper, 10 pt, conference]{ieeeconf}  % Comment this line out if you need a4paper

\IEEEoverridecommandlockouts                              % This command is only needed if 
\usepackage{graphicx} % for pdf, bitmapped graphics files
\usepackage{amsmath} % assumes amsmath package installed
\usepackage{amssymb}  % assumes amsmath package installed
\usepackage{float}
\usepackage{booktabs}
\usepackage{tensor}
\usepackage{accents}
\usepackage{mathtools}
\usepackage{algorithm}
\usepackage{algorithmic}
{
    \newtheorem{assum}{Assumption}
    \newtheorem{thrm}{Theorem}
    
}
\usepackage{wasysym}

\newcommand{\ignore}[1]{}

\newcommand{\bma}[1]{\left[\begin{array}{#1}}
\newcommand{\ema}{\end{array}\right]}

\DeclareMathAlphabet{\mbf}{OT1}{ptm}{b}{n}

\def\fdotb{{\raisebox{-0.6ex}{ \kern0.2ex\raisebox{0.8ex}{\tiny $\hspace*{-1ex}\circ$}}}}
\def\fddotb{{\raisebox{-0.6ex}{ \kern0.2ex\raisebox{0.8ex}{\tiny $\hspace*{-1ex}\circ\circ$}}}}
\newcommand{\utimes}{ {\raisebox{-0.6ex}{ \kern-1.0ex\raisebox{0.6ex}{ \small $\mathsf{v}$}}} } % 
\newcommand{\beq}{\begin{equation}}
\newcommand{\eeq}{\end{equation}}
\newcommand{\bdis}{\begin{displaymath}}
\newcommand{\edis}{\end{displaymath}}
\newcommand{\beqarray}{\begin{eqnarray}}
\newcommand{\eeqarray}{\end{eqnarray}}
\newcommand{\beqarraynn}{\begin{eqnarray*}}
\newcommand{\eeqarraynn}{\end{eqnarray*}}

\DeclareMathOperator*{\argmax}{arg\,max}
\DeclareMathOperator*{\argmin}{arg\,min}

\title{\LARGE \bf Update-Aware Robust Optimal Model Predictive \\ Control for Nonlinear Systems}

\author{Jad Wehbeh$^{1}$ and Eric C. Kerrigan$^{2}$% <-this % stops a space
\thanks{$^{1}$Jad Wehbeh is with the Department of Electrical and Electronic Engineering, Imperial College London, SW7 2AZ, UK
        {\tt\small j.wehbeh22@imperial.ac.uk}}%
\thanks{$^{2}$Eric C. Kerrigan is with the Department of Electrical and Electronic Engineering and the Department of Aeronautics, Imperial College London,
        SW7 2AZ, UK
        {\tt\small e.kerrigan@imperial.ac.uk}}
}

\begin{document}

\maketitle
\thispagestyle{empty}
\pagestyle{empty}

\setlength{\textfloatsep}{5pt}

%%%%%%%%%%%%%%%%%%%%%%%%%%%%%%%%%%%%%%%%%%%%%%%%%%%%%%%%%%%%%%%%%%%%%%%%%%%%%%%%
\begin{abstract}
    Robust optimal or min-max model predictive control (MPC) approaches aim to guarantee constraint satisfaction over a known, bounded uncertainty set while minimizing a worst-case performance bound. Traditionally, these methods compute a trajectory that meets the desired properties over a fixed prediction horizon, apply a portion of the resulting input, and then re-solve the MPC problem using newly obtained measurements at the next time step. However, this approach fails to account for the fact that the control trajectory will be updated in the future, potentially leading to conservative designs. In this paper, we present a novel update-aware robust optimal MPC algorithm for decreasing horizon problems on nonlinear systems that explicitly accounts for future control trajectory updates. This additional insight allows our method to provably expand the feasible solution set and guarantee improved worst-case performance bounds compared to existing techniques. Our approach formulates the trajectory generation problem as a sequence of nested existence-constrained semi-infinite programs (SIPs), which can be efficiently solved using local reduction techniques. To demonstrate its effectiveness, we evaluate our approach on a planar quadrotor problem, where it clearly outperforms an equivalent method that does not account for future updates at the cost of increased computation time.
\end{abstract}

%%%%%%%%%%%%%%%%%%%%%%%%%%%%%%%%%%%%%%%%%%%%%%%%%%%%%%%%%%%%%%%%%%%%%%%

\section{Introduction}
\subsection{Background and Motivation}
Model Predictive Control (MPC) is an optimization-based control approach that has gained widespread adoption in a variety of fields over recent years. MPC exploits knowledge of the system dynamics to provide a feedback-based framework for cost function minimization that enables the handling of a wide range of constraints \cite{rawlings2017model}. In this work, we focus on min-max or robust optimal MPC techniques, which attempt to maximize the performance of the system while guaranteeing constraint satisfaction in the worst case of bounded uncertainties in the model \cite{zagorowska2023automatic}. This allows for a certifiable level of performance to be achieved even when the system dynamics are not known perfectly, and can be highly practical in safety-critical scenarios.

However, while MPC techniques attempt to make use of all of the information available at the time of computing the control trajectory, they generally fail to make use of the knowledge that the control input will be updated at some known point in the future. This can result in suboptimal performance when compared to solutions that utilize this information \cite{scokaert1998min} and may lead to practical problems appearing infeasible despite being solvable, as demonstrated in this work.
On the other hand, dynamic programming approaches to decision making tend to consider all future choices in their optimization process, but struggle to deal with uncertainty and can be cumbersome to use \cite{liu2020adaptive}. This work incorporates dynamic programming approaches into a robust MPC framework to improve performance and reduce conservativeness.

\subsection{State of the Art}

Some attempts have previously been made at leveraging dynamic programming techniques to include information about future control updates in the MPC problem. The works of \cite{scokaert1998min} and \cite{mayne2001control} present some of the earliest examples of such schemes, describing min-max MPC techniques for linear systems that are aware of future feedback. These methods optimize controls over scenario trees for the disturbances which can be constrained to the extremities of the uncertainty due to the linear nature of the system. This was extended to affine systems with polytopic uncertainties and linear constraints in \cite{bjornberg2006approximate}. Other approaches, such as those of \cite{bemporad2003min} (linear systems) and \cite{spjotvold2009inf} (piecewise affine systems), avoid the need for gridding the uncertainty space but compute suboptimal robust solutions. These differ from techniques like those of \cite{kerrigan2004feedback}, compute min-max feedback policies, but forsake the explicit knowledge of future updates to these feedback policies for the sake of computational efficiency. These methods eventually lead to tube-based MPC approaches, which compute feedback policies that ensure all trajectories remain within some invariant tube~\cite{mayne2011tube}. 

More recent examples featuring optimization over feedback policies include \cite{chen2022interactive} which computes different feedback policies for each branch of a finite scenario tree, \cite{xie2021disturbance} which incorporates affine disturbance feedback for nonlinear systems, and \cite{nguyen2024efficient} which incorporates linear matrix inequality constraints to ensure stability. Other relevant advances include the use of min-max differential inequalities to capture the evolution of the uncertainty  in tube-based MPC \cite{villanueva2017robust} and the relaxation of constraints based on the knowledge of feedback to reduce conservativeness \cite{hu2022handling}.  Outside of the MPC literature, we highlight the works of \cite{zagorowska2023automatic} and \cite{wehbeh2024semi} for presenting novel methods based on constraint generation for solving general nonlinear robust optimal control problems.

\subsection{Contributions}
In this paper, we propose a new robust optimal MPC approach for decreasing horizon problems that explicitly accounts for planned future updates to the solution trajectory at known time instances. Decreasing horizon control is a form of model predictive control that keeps a fixed final prediction step across iterations and is ideally suited for finite horizon problems, e.g. batch processes. Unlike existing methods, our solution works for arbitrary nonlinear systems, automatically generates an optimal gridding of the uncertainty space, and returns the true optimal solution to the min-max problem. It does so by casting the problem as a series of nested existence-constrained semi-infinite programs (SIPs) and builds upon existing local reduction methods to compute the uncertainty scenarios. The principal contributions of this work are summarized as follows:

\begin{itemize}
    \item  We derive the novel update-aware {decreasing horizon} robust optimal MPC algorithm and describe its implementation. It is the first solution to the min-max MPC problem with explicit dynamic programming that works for nonlinear systems. It also generates its own optimal uncertainty scenarios, avoiding the issues of complexity and optimality associated with using scenario trees \cite{puschke2018robust}.  
    \item We prove that our algorithm is recursively feasible, guarantees better worst-case performance than methods without update awareness, and ensures improvement in the performance bound between successive iterations.
    \item We validate our proposed approach on a nonlinear planar quadrotor problem, demonstrating the predicted performance improvements over a comparable MPC formulation without update awareness.
\end{itemize}

\section{Problem Description}
{
Let $f_k(\cdot,\cdot,\cdot,\cdot):$ $\mathcal{X} \times \mathcal{X} \times \mathcal{U} \times \mathcal{W} \rightarrow \mathbb{R}^{n_x}$ be the function describing the discrete-time dynamics of the system with state  $x_k$ $\in$ $\mathcal{X}$ $\subseteq$ $\mathbb{R}^{n_x}$ at time-step $k$, associated control inputs $u_k$ $\in$ $\mathcal{U}$ $\subseteq$ $\mathbb{R}^{n_u}$, disturbances $w_k$ $\in$ $\mathcal{W}$ $\subseteq$ $\mathbb{R}^{n_w}$, such that
\begin{equation}
    \label{eq:dyn_standard}
    f_k(x_k,x_{k+1},u_k,w_k) = 0.
\end{equation}
with associated feasible state set
\begin{equation}
    X_k(x_k,u_k,w_k) \coloneqq \left\{ x_{k+1} \left| \,
    f_k(x_k,x_{k+1},u_k,w_k) = 0
    \right. \right\}.
\end{equation}
\textit{Note:} While only explicit control inputs are considered here, it is relatively straightforward to extend $u_k$ to include feedback gains by following the example of \cite{wehbeh2024robust}.

\begin{assum}
\label{assum:compactness_control_uncert}
The sets $\mathcal{U}$ and $\mathcal{W}$ are compact.   
\end{assum}

We consider the problem of controlling this system between time step $0$ and a known final time step $N$ $\in$ $\mathbb{N}$. We denote by $x \in \bar{\mathcal{X}} \coloneq \mathcal{X}^{N+1}$ the collection of states from $k = 0$ to $k = N$, by $u \in \bar{\mathcal{U}} \coloneq\mathcal{U}^N$ the control inputs between~$0$ and $N-1$, and by $w \in \bar{\mathcal{W}} \coloneq \mathcal{W}^{N}$ the uncertainty sequence between~$0$ and $N-1$. The dynamics of~\eqref{eq:dyn_standard} between~$0$ and $N$ can then be combined with an initial condition constraint on $x_0$ into a single function describing the evolution of the state,
\begin{equation}
    \label{eq:dyn_N}
    f(x,u,w) = 0
\end{equation}
where $f(\cdot,\cdot,\cdot):$ $\bar{\mathcal{X}} \times \bar{\mathcal{U}} \times\bar{\mathcal{W}} \rightarrow \mathbb{R}^{N n_x}$. Given any choice of~$u$ and realization of $w$, the arising state trajectory can then be described as
\begin{equation}
\label{eq:feasible_state_set}
X(u,w) \coloneqq \left\{ x \left| \:
    f(x,u,w) = 0
    \right. \right\}.
\end{equation}

\begin{assum}
\label{assum:compactness_state}
$X(u,w)$ is compact $\forall u \in \bar{\mathcal{U}}$, $\forall w \in \bar{\mathcal{W}}$.   
\end{assum}

We then consider $J^*(\cdot):$ $\bar{\mathcal{U}} \rightarrow \mathbb{R}$, the worst case of some cost function $J(\cdot,\cdot,\cdot):$ $\bar{\mathcal{X}} \times \bar{\mathcal{U}} \times \bar{\mathcal{W}} \rightarrow \mathbb{R}$ for a given choice of $u$ over the allowable values of $x$ and $w$, which can be expressed as
\begin{equation}
    J^*(u) \coloneqq \max_{w  \,\in \, \bar{\mathcal{W}}, \, x \, \in \, X(u,w)} J(x,u,w).
\end{equation}

\begin{assum}
\label{assum:bounded_cost}
The cost function $J(\cdot,\cdot,\cdot)$ is bounded $\forall u \in \bar{\mathcal{U}}$, $\forall w \in \bar{\mathcal{W}}$, $\forall x \in X(u,w)$. It follows that the function $J^*(\cdot)$ is similarly bounded $\forall u \in \bar{\mathcal{U}}$.
\end{assum}

We also require that, for our choice of $u$, the system satisfy the inequalities $g(\cdot,\cdot,\cdot):$  $\bar{\mathcal{X}} \times \bar{\mathcal{U}} \times \bar{\mathcal{W}} \rightarrow \mathbb{R}^{n_g}$ under any realization of $w$. Consequently, we can write the set of admissible solutions for $u$ as
\begin{equation}
\mathcal{U}^* \coloneqq \left\{ u \left| \:
    \forall{w} \in \bar{\mathcal{W}}, \, \forall{x} \in X(u,w): \: \: g(x,u,w) \leq 0
    \right. \right\}.
\end{equation}

\begin{assum}
\label{assum:continuity}
The functions $J$ and $g$ are continuous in all of their arguments, i.e., $J(\cdot,\cdot,\cdot), \, g(\cdot,\cdot,\cdot) \in \mathcal{C}^{\,0}$.
\end{assum}

The robust solution to the minimization of $J^*$ under the constraints of $g$ can then be found through the solution of the problem
\begin{equation}
    \label{eq:robust_prob_general}
    \min_{u  \,\in \, \mathcal{U}^*} \; J^*(u)
\end{equation}
which returns the feasible variables $u$ leading to the best worst-case performance in the cost $J$. 

In \cite{wehbeh2024semi}, we showed how the problem of \eqref{eq:robust_prob_general} could be cast as a SIP and an open-loop solution could be obtained through the application of a local reduction method, as described in Section \ref{sec:local_red}. Here, we examine the case where $u$ need not be computed in its entirety at time step 0, and show how updating $u$ within a decreasing horizon context can lead to performance improvements. Through the rest of this paper, we assume that the control inputs can be re-computed at every time step between $0$ and $N$.

 In order to describe our approach, we now introduce $x_{k}^+ \in \mathcal{X}_{k}^+ \coloneqq \mathcal{X}^{N-k}$ to refer to the components of $x$ corresponding to the time steps between $k+1$ and $N$, $u_{k}^+ \in \mathcal{U}_{k}^+ \coloneq \mathcal{U}^{N-k}$ to refer to the decisions of $u$ that are to be made between $k$ and $N-1$. Similarly, we define $w_{k}^+ \in \mathcal{W}_{k}^+ \coloneq \mathcal{W}^{N-k}$ to refer to the time-varying disturbances of $w$ that are realized between $k$ and $N$. We also use the notation $(\cdot)_k^-$ to refer to any quantity between $0$ and $k-1$, with the exception of $x_k^-$, which includes $x_0$ through $x_k$. 

We then introduce $f_k^+(\cdot,\cdot,\cdot,\cdot):$ $\mathcal{X} \times \mathcal{X}_k^+ \times {\mathcal{U}}_k^+ \times {\mathcal{W}}_k^+ \rightarrow \mathbb{R}^{(N-k) n_x}$ to represent the dynamics of the system at time step $k$ given knowledge of $x_k^-$, $u_k^-$, and $\tilde{w}_k^-$, such that
\begin{equation}
    f_k^+(x_{k},x_k^+,u_k^+,w_k^+) = 0
\end{equation}
where $x_k$ is already known. As with \eqref{eq:feasible_state_set}, we define the feasible state trajectory set for each choice of control sequence and uncertainty realization as
    \begin{equation}
    X_{k}^+(x_k,u_{k}^+,w_{k}^+) \coloneqq \left\{ x_{k}^+ \left| \, f_k^+(x_k,x_k^+,u_k^+,w_k^+) = 0
    \right. \right\}.    
    \end{equation}

Next, let $g_{k}(\cdot,\cdot,\cdot,\cdot):$  $\mathcal{Z}_k \times \mathcal{X} \times \mathcal{U} \times \mathcal{W} \rightarrow \mathbb{R}^{n_{g_{k}}}$ refer to any constraints of $g$ featuring elements of $x_{k}$, $u_{k}$, or $w_{k}$, but none from $x_{k+1}^+$, $u_{k+1}^+$, or $w_{k+1}^+$. This is the set of constraints that could not be fully determined before $k$, but are entirely decided after $k+1$. Similarly, let $g_{k}^+(\cdot,\cdot,\cdot,\cdot):$  $\mathcal{Z}_k \times \mathcal{X}_k^+ \times \mathcal{U}_k^+ \times \mathcal{W}_k^+ \rightarrow \mathbb{R}^{n_{g_{k}}^+}$ refer to any constraints containing elements of $x_{k}^+$, $u_{k}^+$, or $w_{k}^+$, which could not be determined before $k$. Here, $\mathcal{Z}_k \coloneq \mathcal{X}_k^- \times \mathcal{U}_k^- \times \mathcal{W}_k^-$ contains all of the information that was determined prior to $k$ and may feature in the constraints, with $z_0 = x_0$.}   

Finally, we define $J_k(\cdot,\cdot,\cdot,\cdot):$ $\mathcal{Z}_k \times \mathcal{X}_k^+ \times \mathcal{U}_k^+ \times \mathcal{W}_k^+ \rightarrow \mathbb{R}$ to be the total cost when $z_k$ is already known such that
\begin{equation}
    J_k(z_k,x_k^+,u_k^+,w_k^+) \coloneqq J(x,u,w)
\end{equation}
and define $J_k^*(\cdot,\cdot):$ $\mathcal{Z}_k \times \mathcal{U}_k^+ \rightarrow \mathbb{R}$ to be the worst-case value of $J_k$ for a given choice of $u_k^+$ such that
\begin{equation}
    J_k^*(z_k,u_k^+) \coloneqq \hspace{-10pt} \max_{\substack{\vphantom{\sum} w_k^+ \in \, \mathcal{W}_{k}^+, \\[1pt] x_{k}^+ \in \, \mathcal{X}_k^+(z_k,u_k^+,w_k^+)}} \hspace{-10pt}J_k(z_k,x_{k}^+,u_k^+,w_k^+).
\end{equation}

If no further updates to $u$ are to occur after time step $k$, and if $u_k^-$ has already been decided, and $w_k^-$ and $x_k^-$ are already determined, the problem of computing the optimal control trajectory reduces back to the formulation described in \cite{wehbeh2024semi}. We refer to the solution over the remaining interval as $\bar{u}_k$, and the associated worst-case cost as $\bar{\gamma}_k$, such that
\begin{subequations}
\begin{align}
    \label{eq:u_bar}
    \bar{u}_k(z_k) & \coloneqq \argmin_{v_k \, \in \, \bar{\mathcal{U}}_k^*(z_k)} \; J_k^*(z_k,v_k)\\
    \label{eq:gamma_bar}
    \bar{\gamma}_k(z_k) & \coloneqq \min_{v_k \, \in \, \bar{\mathcal{U}}_k^*(z_k)} \; J_k^*(z_k,v_k)
\end{align}
\end{subequations}
where $\bar{\mathcal{U}}^*_k(z_k)$ is the set of feasible controls defined by
\begin{equation}
\label{eq:U_bar_star}
\begin{split}
    \bar{\mathcal{U}}^*_k(z_k) \coloneqq & \,\left\{ v_k \in \mathcal{U} \left| \:
    \begin{array}{c} \forall{w}_k^+ \in \mathcal{W}_k^+, \\[1pt] \forall{x}_k^+ \in X_k^+(z_k,u_k^+,w_k^+) \end{array} \, : \right. \right. \\
    & \qquad \qquad \left. \begin{array}{c} \,\\[1pt] \, \end{array} \: \: g_k^+(z_k,x_k^+,u_k^+,w_k^+) \leq 0
     \right\}.
    \end{split}
\end{equation}

Meanwhile, if $k < N-1$ and we know that the controller will be updated in the future, we only need to compute the next segment in the controller trajectory $u^*_{k}(z_k)$, {the control input that, in the worst-case realization of $w_k$, leaves the system in the best possible position to minimize the total cost by the time the control input can next be updated at $k+1$}. $u^*_{k}(z_k)$ also needs to ensure that future constraints remain satisfiable for all possible uncertainties. The rest of the control trajectory, meanwhile, cannot be computed at time $k$ since it relies on information that will only be available in the future. In Section \ref{sec:update_mpc}, we show how to overcome this limitation to propose a novel update-aware robust optimal MPC algorithm that accounts for the acquisition of future information in its decision making process.

\section{Update-Aware Robust Optimal MPC}
\label{sec:update_mpc}

\subsection{Algorithm Description}

In order to derive the update-aware robust optimal MPC algorithm, we begin by applying Bellman's principle of optimality~\cite{bellman1966dynamic} to separate the computation of $u^*_{k}(z_k)$ from the continuation of the trajectory and obtain the recursive problem formulation  
\begin{subequations}
\label{eqs:u_star_bellman}
\begin{equation}
    u^*_{k}(z_k) \coloneqq \argmin_{v_{k} \, \in \, \mathcal{U}} \; \gamma_k
\end{equation}
s.t. $\forall \omega_{{k}} \in \mathcal{W}$, $\forall s_{{k}} \in X_{k}(x_k,v_{k},\omega_{k})$, 
    \begin{align}
        J_{k}^*(z_{k},v_{k}^+) &\leq \gamma_k \\
        (z_k,s_k,v_k,w_k) & = \zeta_{k+1}  \\
        v_{k+1}^+ &= {u^+_{k+1}}^{\hspace{-5pt}*}(\zeta_{k+1}) \\
        g_k(z_k,s_k,v_k,w_k) &\leq 0
    \end{align}
\end{subequations}
where $s$, $v$, $\omega$, and $\zeta$ are the optimization variables associated with $x$, $u$,  $w$, and $z$. As a result of the constraints that are required to hold for all values in the potentially infinite sets $\mathcal{W}$ and $X(z_k,u_k,w_k)$, the problem of \eqref{eqs:u_star_bellman} is classified as a semi-infinite program (SIP). However, due to its recursive nature, \eqref{eqs:u_star_bellman} cannot be solved directly and must instead be converted into an existence-constrained SIP.

In order to do so, we separate the goals of constraint satisfaction and objective minimization and define the set $\mathcal{U}^*_k(\cdot):$ $\mathcal{X} \rightarrow \mathcal{U}$ to be the set of all control input trajectory segments for which the problem remains feasible at $k+1$ for all possible uncertainties. This is expressed as
\begin{equation}
\label{eq:U_star}
\begin{split}
    & \mathcal{U}^*_k(z_k) \coloneqq \\[3pt] &\left\{ v_k \in \mathcal{U}\left|
    \begin{array}{l} \forall{\omega}_k \in \mathcal{W}, \\[1pt] \forall{s}_k \in X_k(z_k,v_k,\omega_k) \end{array} \hspace{-4pt} : \hspace{-1pt} \begin{array}{l} g_k(z_k,s_k,v_k,\omega_k) \leq 0 \\[1pt] \zeta_{k+1} = (z_k,s_k,v_k,\omega_k) \\[1pt]\exists \, v_{k+1} \in \mathcal{U}^*_{k+1}(\zeta_{k+1})
    \end{array} \hspace{-7pt}
     \right. \right\}
    \end{split}
    \raisetag{55pt}
\end{equation}
with $\mathcal{U}^*_N(z_N) = \bar{\mathcal{U}}^*_N(z_N)$.

We also define $\mathcal{J}^*_k(\cdot,\cdot):$ $\mathcal{Z}_k \times \mathbb{R} \rightarrow \mathcal{U}$ as the subset of $\mathcal{U}$ for which there exists a control input trajectory in $\mathcal{U}_{k+1}^+$ leading to a total cost smaller than $\gamma_k$ in the worst case of all possible uncertainties, expressed as
\begin{equation}
\label{eq:J_star}
\begin{split}
    & \mathcal{J}^*_k(z_k,\gamma_k) \coloneqq \,\left\{ v_k \in \mathcal{U} \left| \:
    \begin{array}{l} \forall{\omega}_k \in \mathcal{W}, \\[1pt] \forall{s}_k \in X_k(z_k,v_k,\omega_k) \end{array} \, :\right. \right. \\[1pt]
    &  \; \; \begin{array}{l} \zeta_{k+1} = (z_k,s_k,v_k,\omega_k) \\[3pt] \exists \, v_{k+1} \in \mathcal{U} \left| \:
    \begin{array}{l} \forall{\omega}_{k+1} \in \mathcal{W}, \\[1pt] \forall{s}_{k+1} \in X_{k+1}(\zeta_{k+1},v_{k+1},\omega_{k+1}) \end{array} \, : \right. \end{array} \\
    & \qquad \qquad \qquad \qquad \qquad \quad \vdots \\[1pt]
    &  \; \; \begin{array}{l} \zeta_{N} = (\zeta_{N-1},s_{N-1},v_{N-1},\omega_{N-1}) \\[3pt] \exists \, v_{N} \in \mathcal{U} \left| \:
    \begin{array}{l} \forall{\omega}_{N} \in \mathcal{W}, \\[1pt] \forall{s}_{N} \in X_{N}(\zeta_{N},v_{N},\omega_{N}) \end{array} : \right. \end{array} \\
    & \qquad \qquad \qquad \qquad \left. \: J_{N}(\zeta_{N},s_{N},v_{N},\omega_{N}) \leq \gamma_k \vphantom{\begin{array}{c} \, \\[1pt] \,
    \end{array}} \; 
     \right\}.
    \end{split}
    \raisetag{85pt}
\end{equation}

Problem \eqref{eqs:u_star_bellman} and its optimal objective value are then
\begin{subequations}
\label{eqs:uarompc_star}
\begin{align}
    \label{eq:u_star}
   u^*_{k}(z_k) =& \argmin_{\substack{\delta_k \in \mathbb{R}\\ u_k \in \,{\mathcal{U}_k^*(z_k) \,\cap\, \mathcal{J}^*(z_k,\delta_k)}}} \; \delta_k \\
   \label{eq:gamma_star}
   \gamma^*_{k}(z_k) \coloneqq& \min_{\substack{\delta_k \in \mathbb{R}\\ u_k \in \,{\mathcal{U}_k^*(z_k) \,\cap\, \mathcal{J}^*(z_k,\delta_k)}}} \; \delta_k.
\end{align}
\end{subequations}
The proof for this equivalence relies on \eqref{eq:U_star} and \eqref{eq:J_star} describing the constraints of \eqref{eqs:u_star_bellman} exactly, and is left out of this paper. The update-aware robust optimal MPC algorithm is then built according to a standard MPC structure, using \eqref{eq:u_star} to compute the next section of the control input sequence at every time step, as described in Algorithm \ref{alg:mpc}.

\begin{algorithm}[t]
\begin{algorithmic}[1]
 \caption{\strut Update-Aware Robust Optimal MPC}
 \label{alg:mpc}
 \renewcommand{\algorithmicrequire}{\textbf{Input:}} 
 \renewcommand{\algorithmicensure}{\textbf{Output:}} 
 \REQUIRE $f$, $g$, $J$, $\mathcal{X}$, $\mathcal{U}$, $\mathcal{W}$, $N$, $x_0$.
 \ENSURE  $u^*$, $\gamma^*$
 \\ {\textit{Initialisation} : $z_0 = (x_0)$}
  \FOR {$k = 0$ to $N-1$}
  \STATE Solve \eqref{eqs:uarompc_star} to obtain $u^*_k$ and $\gamma^*_k$ 
  \STATE Apply the control input $u^*_k$
  \STATE Wait until next time step
  \STATE Measure $x_{k+1}$ and $w_k$
  \STATE Set $z_{k+1} = (z_{k},x_{k+1},u^*_k,w_k)$
  \ENDFOR
 \end{algorithmic}
 \end{algorithm}

\subsection{Theoretical Guarantees}
\label{sec:update_mpc_guars}

We now introduce some theoretical results on the performance of the update-aware robust optimal MPC algorithm. 

\begin{thrm}
\label{thrm:U_subset}
Given any $z_k \in \mathcal{Z}_k$, $\bar{\mathcal{U}}^*_k(z_k) \subseteq \mathcal{U}_k^*(z_k)$.
\end{thrm}
\begin{proof}
    Using existential introduction (i.e., $a \in A \implies \exists \, \alpha:\alpha\in A$), we can rewrite \eqref{eq:U_bar_star} recursively as
\begin{equation}
\label{eq:U_bar_star_alt}
\begin{split}
    \bar{\mathcal{U}}^*_k&(z_k) = \left\{ v_k \in \mathcal{U}\left| \: \exists \, v_{k+1} \in \bar{\mathcal{U}}^*_{k+1}(\zeta_{k+1}) \left| \vphantom{\begin{array}{c} \, \\[1pt] \,
    \end{array}}
    \right. \right. \right. \\[3pt]
    & \left. \begin{array}{l} \forall{\omega}_k \in \mathcal{W}, \\[1pt] \forall{s}_k \in X_k(z_k,v_k,\omega_k) \end{array}  : \begin{array}{l} g_k(z_k,s_k,v_k,\omega_k) \leq 0 \\[1pt] \zeta_{k+1} = (z_k,s_k,v_k,\omega_k)
    \end{array} 
     \right\}.
    \end{split}
    \raisetag{48pt}
\end{equation}
    Therefore, $\mathcal{U}_k^*(z_k)$ and $\bar{\mathcal{U}}_k^*(z_k)$ only differ by the ordering of the $\forall$ and $\exists$ operators. Then by universal instantiation, a member of $\bar{\mathcal{U}}_k^*(z_k)$ must also be a member of ${\mathcal{U}}_k^*(z_k)$, since $\exists A_1 \forall A_2 \implies \forall A_2 \exists A_1$ given any $A_1$ and $A_2$. Members of ${\mathcal{U}}_k^*(z_k)$, meanwhile, do not necessarily belong to $\bar{\mathcal{U}}_k^*(z_k)$.
\end{proof}

Theorem \ref{thrm:U_subset} demonstrates that the update-aware method for computing control inputs enables a larger feasible set than similar approaches that neglect this added information. This then leads cleanly to the following result: 
\begin{thrm}
    Given any $z_k \in \mathcal{Z}_k$, $\gamma_k^*(z_k) \leq \bar{\gamma_k}(z_k)$.
\end{thrm}
\begin{proof}
Using existential introduction once again, we can rewrite \eqref{eq:gamma_bar} as
\begin{equation}
\label{eq:gamma_bar_alt}
    \bar{\gamma}_{k}(z_k) = \min_{\substack{\delta_k \in \mathbb{R}\\ u_k \in \,{\bar{\mathcal{U}}_k^*(z_k) \,\cap\, \bar{\mathcal{J}}^*(z_k,\delta_k)}}} \; \delta_k
\end{equation}
where $\bar{\mathcal{J}}^*_k(\cdot,\cdot):$ $\mathcal{Z}_k \times\mathbb{R} \rightarrow \mathcal{U}$ is defined as
\begin{equation}
\begin{split}
    & \bar{\mathcal{J}}^*_k(z_k,\bar{\gamma_k})\coloneqq \left\{ v_k \in \mathcal{U} \left| \, \exists v_{k+1} \in \mathcal{U} \left| \, \hdots \left|  \, \exists v_{N} \in \mathcal{U} \left| \vphantom{\begin{array}{c} \, \\[1pt] \,
    \end{array}} \right. \right. \right. \right. \right. \\[1pt] 
    & \left. 
    \begin{array}{l} \forall{\omega}_{k}^+ \in \mathcal{W}_{k}^+, \\[1pt] \forall{s}_{k}^+ \in X_{k}^+(z_k,v_{k}^+,\omega_{k}^+) \end{array} \hspace{-3pt} : \hspace{-3pt} \right. \left. \: J_k(z_k,s_{k}^+,v_{k}^+,\omega_{k}^+) \leq \gamma_k \vphantom{\begin{array}{c} \, \\[1pt] \,
    \end{array}} \; 
     \right\}.
    \end{split}
    \raisetag{49pt}
\end{equation}
Using the same logic as in Theorem \ref{thrm:U_subset}, we know that $\bar{\mathcal{J}}^*_k(z_k,\bar{\gamma_k}) \subseteq {\mathcal{J}}^*_k(z_k,\bar{\gamma_k})$ $\forall z_k \in \mathcal{Z}_k, \gamma_k \in \mathbb{R}$. It then follows that $\left\{ \bar{\mathcal{U}}_k^*(z_k) \,\cap\, \bar{\mathcal{J}}^*(z_k,\delta_k) \right\} \subseteq \left\{{\mathcal{U}}_k^*(z_k) \,\cap\, {\mathcal{J}}^*(z_k,\delta_k) \right\}$. Therefore, the feasible space for the problem of \eqref{eq:gamma_bar_alt} is contained in that of \eqref{eq:gamma_star}, and given that the two objectives are the same, we have that $\gamma_k(z_1) \leq \bar{\gamma_k}(z_k)$ $\forall z_k \in \mathcal{Z}_k$.
\end{proof}

Consequently, the update-aware robust optimal MPC is guaranteed to lead to a better worst-case performance bound than a traditional robust optimal control method. This does not incur any tradeoff in the feasibility of the solution trajectory computed due to the next theorem we introduce.
\begin{thrm}
    \label{thrm:recursive_feasiblity}
    Algorithm \ref{alg:mpc} is recursively feasible $\forall z_k \in \mathcal{Z}_k$.
\end{thrm}
\begin{proof}
    Recursive feasibility follows trivially from \eqref{eq:U_star} since $u^*_k(z_k) \in \mathcal{U}_k^*(z_k) \implies u_{k+1}^*(z_{k+1}) \in \mathcal{U}_{k+1}^*(z_{k+1})$ $\forall z_k \in \mathcal{Z}_k$, $w_k \in \mathcal{W}$, where $u_k = u_k^*(z_k)$, $x_k = X_k(z_k,u_k,w_k)$, and $z_{k+1} = (z_k,x_k,u_k,w_k)$.
\end{proof}

Finally, we show that the worst-case performance bound $\gamma^*_k$ is guaranteed to decrease between iterations of the update-aware robust optimal MPC algorithm.
\begin{thrm}
    Applying the input $u_k = u_k^*(z_k)$ will always lead to $\gamma^*_{k+1}(z_{k+1}) \leq \gamma^*_k(z_k)$ $\forall z_k \in \mathcal{Z}_k$, $w_k \in \mathcal{W}$, where $x_k = X_k(z_k,u_k,w_k)$, $z_{k+1} = (z_k,x_k,u_k,w_k)$.
\end{thrm}
\begin{proof}
    This follows directly from the knowledge that $u_k \in \mathcal{U}^*(z_k) \,\cap\, \mathcal{J}^*(z_k,\gamma^*_k)$ under \eqref{eq:u_star}. After applying $u_k^*(z_k)$, $\mathcal{J}_{k+1}^*(z_{k+1},\gamma_k^*)$ must be non-empty from \eqref{eq:J_star}. Similarly, $\mathcal{U}^*_{k+1}(z_{k+1})$ is non-empty from Theorem \ref{thrm:recursive_feasiblity}. Therefore, $\gamma_k^*(z_k)$ is guaranteed to be in the solution space of \eqref{eq:gamma_star} for $z_{k+1}$ and the optimal solution can only improve on the previous time-step so that $\gamma^*_{k+1}(z_{k+1}) \leq \gamma^*_k(z_k)$. 
\end{proof}

\section{Solution by Local Reduction}
\label{sec:local_red}

While the problem of \eqref{eq:u_star} gives us the optimal continuation to our control input trajectory, it remains challenging to solve numerically. We address this by noticing that \eqref{eq:u_star} is a nested sequence of $N-k +1$ existence-constrained robust optimal control problems, which, under Assumptions \ref{assum:compactness_control_uncert}--\ref{assum:continuity} can be rewritten as a nested sequence of existence-constrained SIPs using the approach of \cite{wehbeh2024semi}. Then, from \cite{wehbeh2024semi}, \eqref{eqs:uarompc_star} can be reformulated into the sequence of problems
\begin{subequations}
\label{eq:P_k}
\begin{equation}
P_k(z_k) \coloneqq \min_{v_k \in \mathcal{U}, \gamma_k \in \mathbb{R}} \gamma_k
\end{equation}
s.t. $\forall \omega_k \in \mathcal{W}$, $\forall s_k \in X_k(z_k,v_k,\omega_k)$,
\begin{align}
    (z_k,s_k,v_k,\omega_k) & = \zeta_{k+1}\\
    g_k(z_k,x_k,u_k,w_k) & \leq 0 \\
    \label{eq:P_k_sip_constraint}
    P_k^{k+1}(z_k) & \leq \gamma_k
\end{align}
\end{subequations}
with
\begin{subequations}
\label{eq:P_kp1}
\begin{equation}
P_k^{k+1}(\zeta_{k+1}) \coloneqq \min_{v_{k+1} \in \mathcal{U}_{k+1}, \delta_{k+1} \in \mathbb{R}} \delta_{k+1}
\end{equation}
s.t. $\forall \omega_{k+1} \in \mathcal{W}$, $\forall s_{k+1} \in X_{k+1}(\zeta_{k+1},v_{k+1},\omega_{k+1})$,
\begin{align}
    (\zeta_{k+1},s_{k+1},v_{k+1},\omega_{k+1}) & = \zeta_{k+2}\\
    g_{k+1}(\zeta_{k+1},x_{k+1},u_{k+1},w_{k+1}) & \leq 0 \\
    \label{eq:P_kp1_sip_constraint}
    P_k^{k+2}(\zeta_{k+2}) & \leq \delta_{k+1}
\end{align}
\end{subequations}
and
\begin{subequations}
\label{eq:P_N}
\begin{equation}
P_k^{N}(\zeta_{N}) \coloneqq \min_{v_{N} \in \mathcal{U}, \delta_{N} \in \mathbb{R}} \delta_{N}
\end{equation}
s.t. $\forall \omega_{N} \in \mathcal{W}$, $\forall s_{N} \in X_N(\zeta_{N},v_{N},\omega_{N})$,
\begin{align}
    g_{N}(\zeta_{N},x_{N},u_{N},w_{N}) & \leq 0 \\
    J_N(\zeta_N,s_N,v_N,w_N) & \leq \delta_{N}.
\end{align}
\end{subequations}
{where $P_k$ computes the optimal control input $u^*_k$ and $P_{k}^{k+1}$ and $P_k^N$ optimize over feedback policies for the remainder of the trajectory in a dynamic programming sense.}

Each of the problems of \eqref{eq:P_k}, \eqref{eq:P_kp1}, and \eqref{eq:P_N} is then a SIP which can be solved using the local reduction method of \cite{blankenship1976infinitely}. This involves replacing the infinite uncertainty set for each problem by a finite subset of uncertainty scenarios that maximize the violation of the constraints under the current best guess for the decision variables. For \eqref{eq:P_k_sip_constraint} and \eqref{eq:P_kp1_sip_constraint}, this constraint violation maximization, as per \cite{wehbeh2024semi}, requires solving a problem of the form
\begin{subequations}
\label{eqs:lr_max_prob}
\begin{equation}
    w_k^{\text{scen}}(z_k,v_k,\gamma_k) = \argmax_{\substack{\omega_k \in \mathcal{W} \\ s_k \in X_k(z_k,v_k,w_k) \\ \zeta_{k+1} = (z_k,s_k,\omega_k,v_k)}} \sigma_k
\end{equation}
s.t. $\forall \delta_{k+1} \in \mathbb{R}$, $\forall v_{k+1} \in \mathcal{U}_{k+1}^*(\zeta_{k+1}) \cap J^*_{k+1}(\zeta_{k+1},\delta_{k+1})$,
\begin{equation}
    \label{eq:lr_max_prob_gsip_constr}
    \sigma_k - \delta_{k+1} + \gamma_k \leq 0.
\end{equation}
\end{subequations}
{where $\sigma_k \in \mathbb{R}$ is a bound on the constraint violation.}

Unfortunately, due to the dependence of its uncertainty set on a decision variable, \eqref{eqs:lr_max_prob} is a Generalized SIP. We address this using the procedure in \cite{wehbeh2025statedependent}, transforming the constraint of \eqref{eq:lr_max_prob_gsip_constr} into the requirement that $\forall \delta_{k+1} \in \mathbb{R}$, $\forall v_{k+1} \in \mathcal{U}_{k+1}$,
\begin{equation}
\label{eq:lr_max_prob_sip_constr}
\begin{split}
        & (\sigma_k - \delta_{k+1} + \gamma_k \leq 0) \; \lor \\
        & \qquad \qquad\neg \left[v_{k+1} \in\mathcal{U}_{k+1}^*(\zeta_{k+1}) \cap J^*_{k+1}(\zeta_{k+1},\delta_{k+1})\right]
\end{split}
\raisetag{26pt}
\end{equation}
where $\lor$ is the logical or operator and $\neg$ is a negation. The second proposition in \eqref{eq:lr_max_prob_sip_constr} is equivalent to
\begin{equation}
\begin{split}
    & \exists \omega_{k+1}: \neg[{P}_k^{i+2}(\zeta_{k+2}) - \delta_{k+1} \leq 0 ] \; \lor \\ 
     & \qquad\neg [ g_{k+1}(\zeta_{k+1},x_{k+1},u_{k+1},w_{k+1}) \leq 0]
\end{split}
\end{equation}
which, using the smoothing procedure of \cite{wehbeh2025statedependent}, allows us to write \eqref{eq:lr_max_prob_sip_constr} as
\begin{equation}
\label{eq:lr_max_prob_sip_smooth}
\begin{split}
    & \min_{\substack{\omega_{k+1} \in \mathcal{W} \\ \lambda \in \Lambda}} \; \lambda_1 [\sigma_k - \delta_{k+1} +  \gamma_k] \\[-13pt] & \qquad\qquad\quad + \lambda_2 [\delta_{k+1} - \tilde{P}_k^{i+2}(\zeta_{k+2}) + \epsilon] \\
    & \qquad\quad + \lambda_3 [g_{k+1}(\zeta_{k+1},x_{k+1},u_{k+1},w_{k+1})+\epsilon] \leq 0
\end{split}
\end{equation}
where $\tilde{P}_k(z_k) \coloneqq P_k(z_k)$ if $P_k(z_k)$ is feasible and $-\infty$ if not, {$\lambda_1$, $\lambda_2$, and $\lambda_3$ are members of a simplex $\Lambda$, and $\epsilon$ is the smallest positive constant.} Substituting \eqref{eq:lr_max_prob_sip_smooth} for \eqref{eq:lr_max_prob_gsip_constr} in \eqref{eqs:lr_max_prob} then finalizes the conversion into a regular SIP. Notice that solving the constraint generation problem for \eqref{eq:lr_max_prob_sip_smooth} requires solving $P_k^{k+1}(\cdot)$, preserving the recursive structure.

{Since solving each level of the SIP except for $P$ requires solving at least one lower level SIP, we can expect the total number of SIPs that need to be solved to scale linearly with $N$ in the best case (when each SIP converges after a single scenario), and exponentially in the worst case (when all SIPs require multiple scenarios.)} This is consistent with the complexity of similar dynamic programming-based MPC approaches such as those of \cite{scokaert1998min} and \cite{bjornberg2006approximate}. In the case of our approach, however, it is very likely that significant computational improvements can be achieved by exploiting the similarity in the nested problems being solved. Nonetheless, such variations of the proposed method are outside the scope of this paper, and will not be treated further.

{We also remark that the problems \eqref{eq:P_k}, \eqref{eq:P_kp1}, and \eqref{eq:P_N} are usually  non-convex in practice. In this work, we use easily parallelizable multistart methods for scenario generation, leveraging the result from \cite{blankenship1976infinitely} that any constraint-violating scenario can improve the current solution.  If guaranteed optimality is required, global optimization methods may be used, though they come with significant computational cost..} 

\section{Results}

\begin{figure}[b]
    \centering
    \includegraphics[width=0.50\columnwidth]{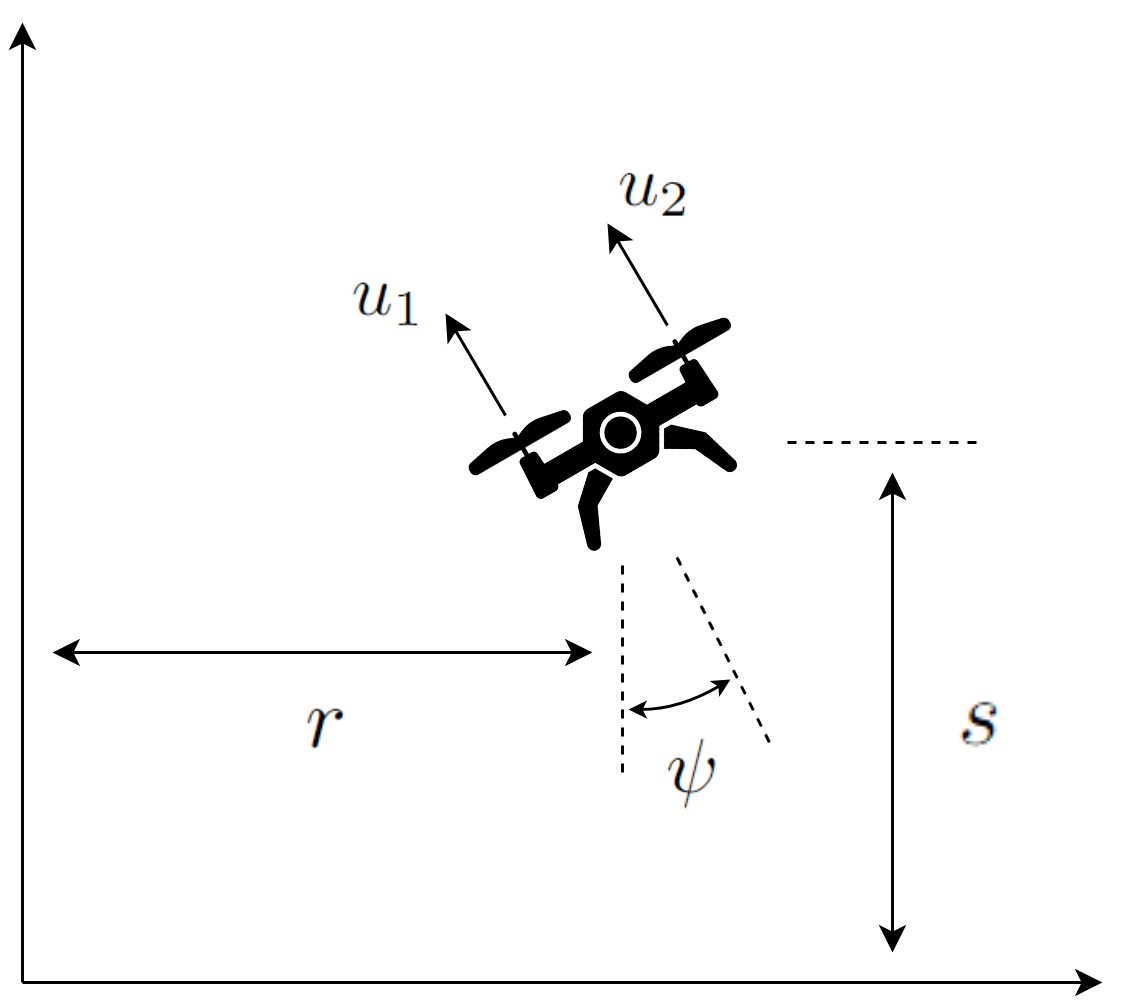}
    \vspace{-5pt}
    \caption{Illustration of the quadrotor's horizontal position ($r$), altitude ($s$), and tilt angle ($\psi$) taken from \cite{wehbeh2024robust}.}
    \label{fig:quad_example}
\end{figure}

In order to test the performance of our proposed update-aware robust optimal MPC algorithm, we compare it to a more standard robust optimal control scheme that applies the control of \eqref{eq:u_bar} instead of \eqref{eq:u_star} on a simulation of a planar quadrotor. The model has continuous-time states $[r,\dot{r},s,\dot{s},\psi,\dot{\psi}]$, where $r$ is the quadrotor's horizontal position, $s$ is the quadrotor's height, and $\psi$ is the quadrotor's tilt angle, as illustrated in Figure \ref{fig:quad_example}.

In continuous-time, the quadrotor's positional and angular dynamics are described by the equations 
\begin{equation}
    \begin{bmatrix}
        \ddot{r}(t) \\ \ddot{s}(t) \\ \ddot{\psi}(t)
    \end{bmatrix}
    =
    \begin{bmatrix}
        \sin(\psi(t)) \left(u_1(t) + u_2(t) \right) /m \\
        \cos(\psi(t)) \left(u_1(t) + u_2(t) \right) /m - b \\
        [\ell (u_1(t) -u_2(t)) + w(t)] / I
    \end{bmatrix}
\end{equation}
where $u_1(t)$ and $u_2(t)$ are the real motor thrusts at time $t$, $m = 0.15$ is the vehicle's mass, $I = 0.00125$ is the moment of inertia, $\ell = 0.1$ is the moment arm for each motor, $b = 9.81$ is the gravity acting on the system, and $w(t)$ is a torque disturbance acting on the system. 

We then obtain the discrete-time dynamics at time step $k$ through the application of a midpoint rule with sampling time $T_s = 0.5$, and where $x^1_{k}$ through $x^6_{k}$ are the discretized states at time $k$ corresponding to the continuous-time values $[r,\dot{r},s,\dot{s},\psi,\dot{\psi}]$ and the discretization is such that $k = 0$ maps to $t = 0$. We examine the behavior of the simulation over 6 time steps, with $\mathcal{K} = \{0,1,2,3,4\}$.

The system is initialized from the origin with the state $x_0 = [0,0,0,0,0,0]$. We also require that the system's $x^1$ coordinate remain between $-c$ and $c$ at all times and that its $x^3$ coordinate remain positive by satisfying the inequalities
\begin{align}
g(x) \coloneqq \left\{  \begin{array}{r} x^1_k - c \\ - x^1_k - c  \\ -x_k^3  \end{array}  \; \forall k \in \{1,\ldots,5\} \right\}  \leq 0  
\end{align}
for all possible uncertainty realizations. The objective of the simulation is to maximize the height achieved by the quadrotor at the final time step such that
\begin{equation}
    J(x) = - x_5^3.
\end{equation}
The system is subject to the input constraints
\begin{equation}
    \mathcal{U} = \left\{ u \left|  \begin{array}{r} -2 \leq  u^1_k \leq 2 \\ -2 \leq  u^2_k \leq 2    \end{array}  \; \forall k \in \{0,\ldots,4\} \right. \right\}
\end{equation}
and the known disturbance set is
\begin{equation}
    \mathcal{W} = \left\{ w \left| -w_{\text{max}} \leq  w_k \leq w_{\text{max}} \; \forall k \in \{0,\ldots,4\} \right. \right\}.
\end{equation}
We compare the two controllers for the scenarios where $c = 1$ and $w_{\text{max}} = 0.001$ (Scenario 1), $c = 0.1$ and $w_{\text{max}} = 0.001$ (Scenario 2), and $c = 1$ and $w_{\text{max}} = 0.01$ (Scenario 3) across 500 simulated runs. The controller of \eqref{eq:u_bar} (RO-MPC) fails to deal with Scenarios 2 and 3, while that of \eqref{eq:u_star} (UARO-MPC) successfully solves all 3 problems. 

\begin{figure}[t]
    \centering
    \includegraphics[width=0.75\columnwidth]{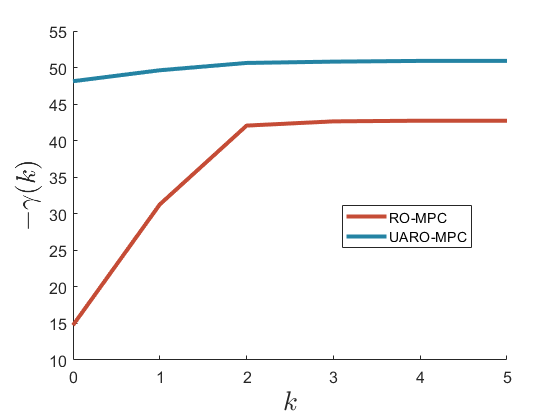}
    \vspace{-10pt}
    \caption{Average $\gamma$ values computed at each time step for Scenario 1.}
    \label{fig:gamma_scen1}
\end{figure}

As seen in Figure \ref{fig:gamma_scen1}, the UARO-MPC algorithm does a much better job of predicting the true performance of the system and achieves a better objective height of 50.91 on Scenario 1 when compared to the RO-MPC's value of 42.73. The UARO-MPC algorithm also performs well on Scenarios 2 and 3, which the RO-MPC approach cannot solve, as seen in Table \ref{tab:perf}. The UARO-MPC approach is also much less conservative in its initial estimate of the performance bound. {This performance improvement, however, comes at the cost of a large increase in computation time.} 

\begin{table}[h]
\caption{\vspace{-2pt} Performance metrics for the two controllers.}
\label{tab:perf}
\vspace{-5pt}
\centering
\begin{tabular}{@{}lcccc@{}}
\toprule
                       & \multicolumn{2}{c}{Scenario 1} & \,Scenario 2 \,& \, Scenario 3 \, \\ \midrule
MPC Type               & RO             & UARO           & UARO       & UARO       \\ \midrule
$-J(\cdot)$            & 42.73          & 50.91          & 48.95      & 42.45      \\ \midrule
$\gamma(0) - J(\cdot)$ & 27.97          & 2.78           & 3.57       & 5.46       \\ \midrule 
{$t_{\text{comp}}$ (s)} & {5.3} & {95.7} &  {101.3} & {104.8}  \\\bottomrule
\end{tabular}
\vspace{-5pt}
\end{table}

\section{Conclusion}

In this paper, we introduced a novel MPC algorithm which extends the idea of using explicit dynamic programming for capturing future control updates to nonlinear systems {by leveraging SIP solution approaches to generate an optimal scenario set}. Our approach is guaranteed to outperform methods that ignore these updates and expands the set of feasible solutions, making previously infeasible problems solvable, as was demonstrated with the planar quadrotor example. While further work is needed to study and optimize the numerical performance of the algorithm, we leave this for future research.

\section*{ACKNOWLEDGMENTS}

This work was funded by the Natural Sciences and Engineering Research Council of Canada through a PGS D grant. 

\addtolength{\textheight}{-0.5cm}   % This command serves to balance the column lengths

%%%%%%%%%%%%%%%%%%%%%%%%%%%%%%%%%%%%%%%%%%%%%%%%%%%%%%%%%%%%%%%%%%%%%%%%%%%%%%%%

%%%%%%%%%%%%%%%%%%%%%%%%%%%%%%%%%%%%%%%%%%%%%%%%%%%%%%%%%%%%%%%%%%%%%%%%%%%%%%%%
%%  Bibliography
%%%%%%%%%%%%%%%%%%%%%%%%%%%%%%%%%%%%%%%%%%%%%%%%%%%%%%%%%%%%%%%%%%%%%%%%%%%%%%%%

\bibliographystyle{IEEEtran}
\bibliography{IEEEabrv,references.bib}

\begin{thebibliography}{10}
\providecommand{\url}[1]{#1}
\csname url@samestyle\endcsname
\providecommand{\newblock}{\relax}
\providecommand{\bibinfo}[2]{#2}
\providecommand{\BIBentrySTDinterwordspacing}{\spaceskip=0pt\relax}
\providecommand{\BIBentryALTinterwordstretchfactor}{4}
\providecommand{\BIBentryALTinterwordspacing}{\spaceskip=\fontdimen2\font plus
\BIBentryALTinterwordstretchfactor\fontdimen3\font minus \fontdimen4\font\relax}
\providecommand{\BIBforeignlanguage}[2]{{%
\expandafter\ifx\csname l@#1\endcsname\relax
\typeout{** WARNING: IEEEtran.bst: No hyphenation pattern has been}%
\typeout{** loaded for the language `#1'. Using the pattern for}%
\typeout{** the default language instead.}%
\else
\language=\csname l@#1\endcsname
\fi
#2}}
\providecommand{\BIBdecl}{\relax}
\BIBdecl

\bibitem{rawlings2017model}
J.~B. Rawlings, D.~Q. Mayne, M.~Diehl \emph{et~al.}, \emph{Model predictive control: theory, computation, and design}.\hskip 1em plus 0.5em minus 0.4em\relax Nob Hill Publishing Madison, WI, 2017, vol.~2.

\bibitem{zagorowska2023automatic}
M.~Zagorowska, P.~Falugi, E.~O'Dwyer, and E.~C. Kerrigan, ``Automatic scenario generation for efficient solution of robust optimal control problems,'' \emph{International Journal of Robust and Nonlinear Control}, vol.~34, no.~2, pp. 1370--1396, 2024.

\bibitem{scokaert1998min}
P.~O. Scokaert and D.~Q. Mayne, ``Min-max feedback model predictive control for constrained linear systems,'' \emph{IEEE Transactions on Automatic control}, vol.~43, no.~8, pp. 1136--1142, 1998.

\bibitem{liu2020adaptive}
D.~Liu, S.~Xue, B.~Zhao, B.~Luo, and Q.~Wei, ``Adaptive dynamic programming for control: A survey and recent advances,'' \emph{IEEE Transactions on Systems, Man, and Cybernetics: Systems}, vol.~51, no.~1, pp. 142--160, 2020.

\bibitem{mayne2001control}
D.~Q. Mayne, ``Control of constrained dynamic systems,'' \emph{European Journal of Control}, vol.~7, no. 2-3, pp. 87--99, 2001.

\bibitem{bjornberg2006approximate}
J.~Bj{\"o}rnberg and M.~Diehl, ``Approximate robust dynamic programming and robustly stable {MPC},'' \emph{Automatica}, vol.~42, no.~5, pp. 777--782, 2006.

\bibitem{bemporad2003min}
A.~Bemporad, F.~Borrelli, and M.~Morari, ``Min-max control of constrained uncertain discrete-time linear systems,'' \emph{IEEE Transactions on automatic control}, vol.~48, no.~9, pp. 1600--1606, 2003.

\bibitem{spjotvold2009inf}
J.~Spj{\o}tvold, E.~Kerrigan, D.~Mayne, and T.~Johansen, ``Inf--sup control of discontinuous piecewise affine systems,'' \emph{International Journal of Robust and Nonlinear Control: IFAC-Affiliated Journal}, vol.~19, no.~13, pp. 1471--1492, 2009.

\bibitem{kerrigan2004feedback}
E.~C. Kerrigan and J.~M. Maciejowski, ``Feedback min-max model predictive control using a single linear program: robust stability and the explicit solution,'' \emph{International Journal of Robust and Nonlinear Control: IFAC-Affiliated Journal}, vol.~14, no.~4, pp. 395--413, 2004.

\bibitem{mayne2011tube}
D.~Q. Mayne, E.~C. Kerrigan, E.~Van~Wyk, and P.~Falugi, ``Tube-based robust nonlinear model predictive control,'' \emph{International journal of robust and nonlinear control}, vol.~21, no.~11, pp. 1341--1353, 2011.

\bibitem{chen2022interactive}
Y.~Chen, U.~Rosolia, W.~Ubellacker, N.~Csomay-Shanklin, and A.~D. Ames, ``Interactive multi-modal motion planning with branch model predictive control,'' \emph{IEEE Robotics and Automation Letters}, vol.~7, no.~2, pp. 5365--5372, 2022.

\bibitem{xie2021disturbance}
H.~Xie, L.~Dai, Y.~Lu, and Y.~Xia, ``Disturbance rejection {MPC} framework for input-affine nonlinear systems,'' \emph{IEEE Transactions on Automatic Control}, vol.~67, no.~12, pp. 6595--6610, 2021.

\bibitem{nguyen2024efficient}
T.~H. Nguyen, D.~Q. Bui, P.~N. Dao \emph{et~al.}, ``An efficient min/max robust model predictive control for nonlinear discrete-time systems with dynamic disturbance,'' \emph{Chaos, Solitons \& Fractals}, vol. 180, p. 114551, 2024.

\bibitem{villanueva2017robust}
M.~E. Villanueva, R.~Quirynen, M.~Diehl, B.~Chachuat, and B.~Houska, ``Robust {MPC} via min--max differential inequalities,'' \emph{Automatica}, vol.~77, pp. 311--321, 2017.

\bibitem{hu2022handling}
J.~Hu, X.~Lv, H.~Pan, and M.~Zhang, ``Handling the constraints in min-max {MPC},'' \emph{IEEE Transactions on Automation Science and Engineering}, vol.~21, no.~1, pp. 296--304, 2022.

\bibitem{wehbeh2024semi}
J.~Wehbeh and E.~C. Kerrigan, ``Semi-infinite programs for robust control and optimization: Efficient solutions and extensions to existence constraints,'' in \emph{8th IFAC Conference on Nonlinear Model Predictive Control NMPC 2024}.\hskip 1em plus 0.5em minus 0.4em\relax IFAC, 2024, pp. 317--322.

\bibitem{puschke2018robust}
J.~Puschke, H.~Djelassi, J.~Kleinekorte, R.~Hannemann-Tam{\'a}s, and A.~Mitsos, ``Robust dynamic optimization of batch processes under parametric uncertainty: Utilizing approaches from semi-infinite programs,'' \emph{Computers \& Chemical Engineering}, vol. 116, pp. 253--267, 2018.

\bibitem{wehbeh2024robust}
J.~Wehbeh and E.~C. Kerrigan, ``Robust output feedback of nonlinear systems through the efficient solution of min-max optimization problems,'' in \emph{2024 IEEE 63rd Conference on Decision and Control (CDC)}.\hskip 1em plus 0.5em minus 0.4em\relax IEEE, 2024, pp. 8870--8875.

\bibitem{bellman1966dynamic}
R.~Bellman, ``Dynamic programming,'' \emph{science}, vol. 153, no. 3731, pp. 34--37, 1966.

\bibitem{blankenship1976infinitely}
J.~W. Blankenship and J.~E. Falk, ``Infinitely constrained optimization problems,'' \emph{Journal of Optimization Theory and Applications}, vol.~19, pp. 261--281, 1976.

\bibitem{wehbeh2025statedependent}
\BIBentryALTinterwordspacing
J.~Wehbeh and E.~C. Kerrigan, ``State-dependent uncertainty modeling in robust optimal control problems through generalized semi-infinite programming,'' 2025. [Online]. Available: \url{https://arxiv.org/abs/2503.10389}
\BIBentrySTDinterwordspacing

\end{thebibliography}

\end{document}